\documentstyle[preprint,aps,epsf]{revtex}

\newcommand{\be}{\begin{eqnarray}}
\newcommand{\ee}{\end{eqnarray}}
\newcommand{\la}{\langle}
\newcommand{\ra}{\rangle}

\newcommand{\eps}{\epsilon}

\newcommand{\Dmatrix}[4]{
        \left(
        \begin{array}{cc}
        #1  & #2   \\
        #3  & #4   \\
        \end{array}
        \right)
        }

\begin{document}
%\twocolumn
\title
{Dual kinetic balance
 approach to  basis set expansions for the Dirac equation}
\author{V. M. Shabaev$^{1,2}$, 
I. I. Tupitsyn$^{1}$, V. A. Yerokhin$^{1,2}$, G. Plunien$^{2}$,
and G. Soff$^{2}$}

\address
{$^{1}$Department of Physics, St.Petersburg State University, Oulianovskaya 1,
Petrodvorets, St.Petersburg 198504, Russia\\
$^{2}$
Institut f\"ur Theoretische Physik, TU Dresden, Mommsenstrasse 13,
D-01062 Dresden, Germany}
%\date
\maketitle
%\onecolumn
\begin{abstract}
A new approach to finite basis sets for the Dirac equation is developed.
It  solves the problem of spurious states and,
as a result, improves the convergence properties of basis set calculations.
The efficiency of the method
is demonstrated for finite basis
sets constructed from B splines by calculating the one-loop 
 self-energy correction for a hydrogenlike ion.  

\end{abstract}
%\twocolumn
%\newpage

%\twocolumn

At present, a great variety of calculations in atomic physics and
quantum chemistry are based on finite basis sets.
First successful attempts  to utilize
 finite basis sets
in relativistic quantum mechanics were made many years ago
\cite{kim67,brat77,kag80,dra81,gra82,kut84}. 
Application of B splines for constructing basis sets 
\cite{sch85,joh86,joh88} provided
 new impact to this field.
Nowdays, B splines are widely employed
 in computational atomic
and molecular physics (see \cite{sap96,bac01} and references therein).

In contrast to
the nonrelativistic case, the use of finite basis sets 
in the
relativistic theory is generally accompanied by the occurrence
of spurious states (see, e.g., \cite{dra81,joh88}). 
For the attractive  Coulomb potential, spurious states appear
for $\kappa>0$ as 
 the lowest bound states with non-physical energies
($\kappa=(-1)^{j+l+1/2}(j+1/2)$ is the quantum number determined
by the angular momentum and the parity of the state).
 For the point nucleus, their 
energies 
coincide with the physical energies of the lowest
bound states with $\kappa<0$.
The wave functions of these
states oscillate rapidly and, therefore,
 in many cases they may be disregarded
in practical atomic calculations.
 However, since the presence of
the spurious states disturbs the spectrum, 
it worsens the convergence properties of the basis set
calculations in some cases.
For this reason,
despite of the early promising results \cite{blu91,blu93},
the finite basis set method has not been extensively employed
in calculations of radiative corrections. 
To date, most of these calculations are
performed by means of
 analytical or numerical representations
for the Coulomb-Green function  
(see, e.g., \cite{moh93,art97,yer97,sap02}) 
or by
the space discretization method \cite{sal89,per97,sun98}, 
in which the  spurious states
are eliminated from the very beginning
(see \cite{sal89} for details).
The problem of spurious states is especially demanding
 in calculations
of atoms in strong external fields and for molecules, where,
generally speaking,  they can not be selected as the lowest bound-state
energies and, therefore, can not be eliminated without employing 
special methods.
Furthermore, even in those cases where 
spurious states do not play any significant role,
their presence generates  
difficulties for providing
an adequate estimate  of the accuracy of 
the calculations.
 
A number of proposals for solving the problem of spurious
states were presented previously  
\cite{gol85,joh88,qui87,gra98,dol00}. Our analysis of
these methods indicates, however, that none of them can be 
considered as completely satisfactory.  
We find that treating the problem
 by a particular choice of the 
boundary conditions in the B-spline method \cite{joh88} 
does not provide any improvement in 
calculations of radiative corrections.
The other methods either
require considerable modifications of the standard 
numerical procedure or they are strongly limited to a specific
choice of finite basis sets. Their applicability 
to calculations
 of radiative corrections thus remains 
 questionable.
For this reason,  we suggest a new method for
solving the problem of spurious states. The efficiency
of the method is demonstrated for finite basis sets
constructed from B splines by calculating
the one-loop self-energy correction for a hydrogenlike ion. 

For the case of a central field $V(r)$,
the Dirac wave function is conveniently represented by
\begin{eqnarray}
\psi({\bf r})=
\frac{1}{r}\left(\begin{array}{c}
G(r)\;\Omega_{\kappa m}({\bf n})\\
iF(r)\;\Omega_{-\kappa m}({\bf n})
\end{array}\right)\;,
\end{eqnarray}
where
${\bf n}={\bf r}/r$. With this representation, the radial Dirac 
equations can be written as
\begin{eqnarray}
H_{\kappa}\phi=E\phi\,,
\end{eqnarray}
where (in units: $\hbar=1$)
\begin{eqnarray}
H_{\kappa}=\Dmatrix{mc^2+V}
{\;\;\displaystyle{c\Bigl[-\frac{d}{dr}+\frac{\kappa}{r}\Bigr]}}
{\displaystyle{c\Bigl[\frac{d}{dr}+\frac{\kappa}{r}\Bigr]}}{\;\;-mc^2+V}
\end{eqnarray}
and
\begin{eqnarray}
\phi({ r})=
\left(\begin{array}{c}
G(r)\\
F(r)
\end{array}\right)\;
\end{eqnarray}
is the two-component radial wave function.
The scalar product of the two-component functions is defined by 
\begin{eqnarray}
\langle \phi_a|\phi_b\rangle
=\int_{0}^{\infty}dr\,[G_a(r)G_b(r)+F_a(r)F_b(r)]\,.
\end{eqnarray}
The radial Dirac equations can be derived from an action principle
$\delta S =0$ with
\begin{eqnarray}
S=\la \phi|H_{\kappa}|\phi\ra -E\la \phi |\phi \ra\,,
\end{eqnarray}
if proper boundary conditions for $G(r)$ and $F(r)$
are implemented.
The functions $\phi(r)$ can be approximated by 
\begin{eqnarray}
\phi(r)=\sum_{i=1}^{2n}c_i u_i(r)\,,
\end{eqnarray}
where the two-component functions $u_i(r)$ are assumed to 
be square integrable, linearly independent, and satisfying 
proper boundary condition at $r=0$.
The variational principle
reduces to the following algebraic equations
\begin{eqnarray}
dS/dc_i=0\,,\;\;\;\;i=1,2,...,2n\,.
\end{eqnarray}
This leads to a generalized eigenvalue problem
\begin{eqnarray} \label{geneig}
K_{ik}c_k=EB_{ik}c_k\,,
\end{eqnarray}
where $K_{ik}=(\la u_i|H_{\kappa}|u_k\ra +
\la u_k|H_{\kappa}|u_i\ra)/2$,
$B_{ik}=\la u_i|u_k\ra$, and the summation over repeated indices
is implied. 

Let us first demonstrate that the widely applied choice
\begin{eqnarray} \label{u1}
u_i({ r})&=&\left(\begin{array}{c} \pi_i(r)\\
0
\end{array}\right)\,,\; i=1,...,n\,,\\
u_i({ r})&=&\left(\begin{array}{c} 0\\
\pi_{i-n}(r)
\end{array}\right)\,,\; i=n+1,...,2n\,,\,
\label{u2}
\end{eqnarray}
where  $\{\pi_i(r)\}_{i=1}^n$  are square integrable functions
satisfying the boundary condition $\pi_i(0)=0$,
 results in the occurrence of spurious states.
In this case, equation (\ref{geneig}) takes the form
\begin{eqnarray}
\label{mat1}
(mc^2+V-E)_{ik}p_k+cD_{ik}q_k &=& 0 \,, \\
c(D^{\dag})_{ik}p_k+(-mc^2+V-E)_{ik}q_k &=& 0\,, 
\label{mat2}
\end{eqnarray}
where $(\pm mc^2+V-E)$, 
% $(-mc^2+V-E)$, 
$D$, and $D^{\dag}$
are $n\times n$ matrices with elements
\begin{eqnarray}
%&&
(\pm mc^2+V-E)_{ik}
% \;\;\;\;\;\;\;\;\;\;\;\; \;\;\;\;\;\;\;\;\;\;\;\; \;\;\;\;\;\;\;\;\;\;\;\; 
%\\ \nonumber
%&&
&=&\int_0^{\infty} dr\;\pi_i(r) (\pm mc^2+V-E)\pi_k(r)\,,\\
%\end{eqnarray}
%\begin{eqnarray}
D_{ik}&=&\int_0^{\infty} dr\;\pi_i(r) \Bigl(-\frac{d}{dr}+
\frac{\kappa}{r}\Bigr)\pi_k(r)\,,\\
(D^{\dag})_{ik}&=&\int_0^{\infty} dr\;\pi_i(r) \Bigl(\frac{d}{dr}+
\frac{\kappa}{r}\Bigr)\pi_k(r)\,,
\end{eqnarray}
and $p_i=c_i$, $q_i=c_{i+n}$ for $i=1,2,...,n$.
Let us consider the nonrelativistic limit
($c \rightarrow \infty$) and introduce vectors 
 $P$ and $Q$ with components
$\{p_i\}_{i=1}^n$ and $\{q_i\}_{i=1}^n$, respectively. 
Then Eq. (\ref{mat2}) yields $Q=(1/2mc)D^{\dag}P$.
Substituting this expression into Eq. (\ref{mat1}),
we obtain
\begin {eqnarray}
DD^{\dag}P+2m(mc^2+V-E)P=0\,.
\label{mat3}
\end{eqnarray}
For the pure Coulomb field, $V(r)=-\lambda/r$ ($\lambda>0$),
introducing the matrix $C_{ik}=D_{ik}-(m\lambda/\kappa) \delta_{ik}$,
Eq. (\ref{mat3}) reduces to
\begin {eqnarray}
C_{\kappa}C_{\kappa}^{\dag}P=\eps P\,,
\label{mat4}
\end{eqnarray}
where $\eps=[2m(E-mc^2)+m^2\lambda^2/\kappa^2]$ and the
dependence of the $C$ matrix on $\kappa$ is explicitly 
indicated.
Taking into account that 
$C_{-\kappa}=-C_{\kappa}^{\dag}$,
we find that the corresponding equation for $\kappa'=-\kappa$
can be written as
\begin {eqnarray}
C_{\kappa}^{\dag}C_{\kappa}P'=\eps' P'\,.
\label{mat5}
\end{eqnarray}
Multiplying Eq. (\ref{mat4}) with 
$C_{\kappa}^{\dag}$, we obtain that each nonzero eigenvalue
of $C_{\kappa}C^{\dag}_{\kappa}$ is an eigenvalue of
$C_{\kappa}^{\dag}C_{\kappa}$. Evidently, the inverse 
statement can be proven in a similar manner and the dimension
of a nonzero eigenvalue subspace is the same for
$C_{\kappa}C^{\dag}_{\kappa}$ and $C_{\kappa}^{\dag}C_{\kappa}$.
Accordingly, the spectra of $C_{\kappa}C^{\dag}_{\kappa}$ and
$C_{\kappa}^{\dag}C_{\kappa}$ may differ only by the dimension
of the zero eigenvalue subspace. For finite matrices, the
dimension of the subspace with $\eps=0$  is the same for
$C_{\kappa}C^{\dag}_{\kappa}$ and
$C_{\kappa}^{\dag}C_{\kappa}$, since the total number of
eigenvectors as well as the dimension of the nonzero eigenvalues
subspace is the same for $C_{\kappa}C^{\dag}_{\kappa}$
and $C_{\kappa}^{\dag}C_{\kappa}$. 
Therefore, the finite matrices $C_{\kappa}C^{\dag}_{\kappa}$
and $C_{\kappa}^{\dag}C_{\kappa}=C_{-\kappa}C^{\dag}_{-\kappa}$
have an identical spectrum. Conversely, we know
that the exact analytical solution of the Dirac equation
for the Coulomb potential yields different lowest bound-state
energies for $\kappa<0$ and $\kappa>0$. This is due to 
the fact that within the exact (infinite dimension) treatment
the subspace with $\eps=0$ may have different dimensions for 
$\kappa<0$ and $\kappa>0$ cases. This can easily be checked by
solving the equation 
\begin{eqnarray}\label{spur}
(d/dr+\kappa/r-m\lambda/\kappa)G(r)=0\,,
\end{eqnarray}
which in case of finite dimensions is equivalent to
the equation $C_{\kappa}^{\dag}P=0$. Solving  equation (\ref{spur})
yields
$G(r)=A_0 r^{-\kappa} \exp{(m\lambda/\kappa r)}$. For $\kappa<0$,
this solution has the proper behaviour at $r\rightarrow 0$
and at $r\rightarrow \infty$. However, this does not hold 
for $\kappa>0$.
Concluding, in the approximation of finite dimensions,
 our proof clearly
indicates the  presence of 
spurious states 
with energy $E-mc^2=-m\lambda^2/2\kappa^2$  corresponding
to $\eps=0$.
It can be shown that this result remains valid for
the full relativistic theory as well. It is obvious that spurious
states must occur for any other potential one is dealing with 
in atomic calculations.

To eliminate the spurious states, in Ref. \cite{qui87}
"kinetically balanced" Slater type  functions
were employed. Within this method,  
for $\kappa>0$ the lower components 
 in equation (\ref{u2}) 
are replaced by functions $\rho_i(r)$ which,
in the nonrelativistic limit, are related to the 
upper components  $\pi_i(r)$
in equation (\ref{u1})
via
\begin{eqnarray}
\rho_i(r)\approx (1/2mc)(d/dr+\kappa/r)\pi_i(r)\,.
\end{eqnarray}
This method provides a high accuracy in calculations
of bound-state energies in atoms
for both sign of $\kappa$, if
an extended nuclear charge distribution is introduced
and proper boundary conditions are implemented 
\cite{qui87}.
However, since the basis functions are
"kinetically balanced" for positive energy states only, 
the application of this 
method to calculations of the QED corrections
might be problematic \cite{qui94}.
The equivalent treatment of  positive and negative 
energy states would provide, in particular,
the well-known symmetry properties of 
the Dirac spectrum under 
the transformations
$V\rightarrow -V$, $\kappa \rightarrow -\kappa$,
and  $G \leftrightarrow F$.
 It is evident that the "kinetically balanced"
functions do not meet with this requirement.

In the original version of the B-spline method 
\cite{joh86,joh88,sap96}, 
to achieve that
 the first positive-energy states $\kappa>0$
correspond to physical bound states,
an additional term had to be
introduced in the Hamiltonian,
which formally implements the so-called MIT boundary condition 
\cite{mit}:
$G(R)=F(R)$, where $R$ is the cavity radius, together with  
the condition $G(0)=0$. 
However, since the presence of the additional term
does not imply any practical advantages, 
it is usually omitted in calculations.
Instead, the boundary conditions are generally
implemented
by eliminating the first and the last basis 
function which are the only ones that do not vanish
at $r=0$ and $r=R$, respectively.
This method was successfully employed
for calculations of the
two-photon exchange diagrams within the rigorous QED approach
\cite{blu93b,yer00,moh00,and01,art03} 
and for relativistic calculations
of the recoil effect \cite{art95,sha98,sha02}.
 However, its application
 to calculations of pure radiative corrections
\cite{blu91,blu93} was less successful, compared to the other methods
\cite{moh93,art97,yer97,sap02,per97,sun98}.
We conjecture
that this would not be the case if the spurious
states were eliminated  in a more natural manner than 
it was done in \cite{joh86,joh88}.

It is known (see, e.g., \cite{blu93,sap96,qui87})
that the case of the pure  Coulomb potential
requires generally a special care in implementing finite
basis set methods. This is due to the singularity of the
Coulomb potential at $r\rightarrow 0$. However,
in practical calculations it is standard to modify
the potential to account for the finite nuclear size, 
which eliminates this problem. For this reason and for
simplicity, we restrict our consideration
to the case of a finite nuclear charge distribution,
bearing in mind that the limit of
a point nucleus  can be treated
by extrapolating a series
 of calculations for
extended nuclei
to vanishing nuclear size.
For extended nuclei
we propose to employ the following basis set
\begin{eqnarray} \label{u1new}
u_i({ r})&=&\left(\begin{array}{c} 
\displaystyle{\pi_i(r)}\\
\rule{0pt}{6mm}
\displaystyle{\frac{1}{2mc}\Bigl(\frac{d}{dr}
+\frac{\kappa}{r}\Bigr)\pi_i(r)}
\end{array}\right)\;, i\le n\,,\\
\rule{0pt}{9mm}
u_i({ r})&=&\left(\begin{array}{c} 
\displaystyle{\frac{1}{2mc}\Bigl(\frac{d}{dr}
-\frac{\kappa}{r}\Bigr)\pi_{i-n}(r)}\\
\rule{0pt}{5mm}
 \displaystyle{\pi_{i-n}(r)}
\end{array}\right)\;,  i \ge n+1\,,
\label{u2new}
\end{eqnarray}
where the linearly independent
functions $\{\pi_i(r)\}_{i=1}^{n}$ are assumed to 
be square integrable and satisfying the proper boundary
condition at $r=0$.
We state that this basis set 
satisfies the following
requirements:
\begin{enumerate}
\item{It is symmetric with respect to the replacement
$\kappa \rightarrow -\kappa$ and the interchange of the upper and
lower components.}
\item{The functions $u_1$, ..., $u_n$ provide the 
correct relation between upper and lower
 components
 for $|E-mc^2|,\, |V(r)|\ll 2mc^2$, while
the functions $u_{n+1}$, ..., $u_{2n}$ do the same
 for $|E+mc^2|,\,|V(r)|\ll 2mc^2$.} 
\item{Calculations utilizing
the standard finite 
basis set determined by Eqs. (\ref{u1}) and (\ref{u2}) 
can be easily adopted when employing the basis
(\ref{u1new})-(\ref{u2new}).}
\item{No spurious states occur for attractive
as well as for repulsive potentials.}
\end{enumerate}
The properties 1 - 3  follow immediately
from definitions 
(\ref{u1new}) and (\ref{u2new}). 
The absence of spurious states 
can be explained
as follows. Performing similar steps as for the derivation
of  Eq. (\ref{mat3}),
for $|E-mc^2|\ll 2mc^2$
we obtain
\begin {eqnarray}
\frac{1}{2m} \, LP \,+\, (V+mc^2-E)P=0\,,
\label{mat3new}
\end{eqnarray}
where 
\begin{eqnarray}
L_{ik}=\int_0^{\infty} dr\;\pi_i(r) \Bigl(-\frac{d}{dr}+
\frac{\kappa}{r}\Bigr) \Bigl(\frac{d}{dr}+
\frac{\kappa}{r}\Bigr)
\pi_k(r)\,.
\end{eqnarray}
Eq. (\ref{mat3new}) takes
 the form of the ordinary Schr\"odinger equation
with $l=|\kappa|+(\kappa/|\kappa|-1)/2$
in the finite basis representation. As is known,
it generates no spurious states. The region $|E+mc^2|\ll 2mc^2$,
where spurious states may exist for repulsive potentials and
for $\kappa<0$, can be considered similarly.
In this case, we obtain the equation 
\begin {eqnarray}
\frac{1}{2m} \, MQ \,+\, (-V+mc^2+E)Q=0\,,
\label{mat4new}
\end{eqnarray}
where 
\begin{eqnarray}
M_{ik}=\int_0^{\infty} dr\;\pi_i(r) \Bigl(\frac{d}{dr}+
\frac{\kappa}{r}\Bigr) \Bigl(-\frac{d}{dr}+
\frac{\kappa}{r}\Bigr)
\pi_k(r)\,.
\end{eqnarray}
Eq. (\ref{mat4new}) has also the form of the ordinary
Schr\"odinger equation but with 
 $l'=|\kappa|-(\kappa/|\kappa|+1)/2$. 
It transforms into 
 equation (\ref{mat3new}) under the replacements
$\kappa\rightarrow -\kappa$, $V\rightarrow -V$,
 $E\rightarrow -E$, $Q\rightarrow P$ and
does not generate
any spurious states. 
This is a consequence
of the equivalent treatment
of the positive and negative energy states.
For this reason, the new basis may be termed
conventionally as  dual kinetic-balance (DKB) basis.

The validity of statement 4 has also been proven by  numerical
calculations with $\pi_i(r)=B_i(r)$, where $B_i(r)$ are the
B splines defined on the interval $(0,R)$ 
as in Ref. \cite{joh88}. The first  and the last spline function
 have been 
omitted. 
 Standard test calculations (see, e.g., \cite{joh88})
show that this basis satisfies suitable
completeness criteria as $n\rightarrow \infty$.

Finally, let us consider the calculation of the one-loop
self-energy (SE) correction to the ground-state energy of 
a hydrogenlike ion employing the new basis set.
Generally, the SE correction
is expanded 
into the zero-, one-, and many-potential terms. 
The ultraviolet divergences in the zero- and one-potential
terms and in the counterterm cancel each other
and their evaluation
can be performed
according to the
formulas presented in Refs. \cite{sny91,yer99}.
As to the many-potential term, although it does not
contain any ultraviolet divergences, its calculation
is most difficult since it involves the summation over
the whole Dirac-Coulomb spectrum. 
In Table 1, we compare our results 
obtained for the many-potential term 
for  $Z=20$ employing the DKB basis set
(\ref{u1new}), (\ref{u2new})
with $\pi_i(r)=B_i(r)$, the old basis
(\ref{u1}), (\ref{u2})  with the same $\pi_i(r)$, and
the results of
a calculation using the analytical representation
for the Coulomb-Green function. The shell model for the nuclear
charge distribution has been used with the radius $R=3.478$ fm.
In the basis set calculations,
the contributions with $|\kappa|\ge 10$
were obtained by an extrapolation. 
Adding the zero- and one-potential terms 
to the many-potential term
yields 
 0.06409 a.u. for the old basis,
 0.06426 a.u. for the DKB basis,
and 0.06425(1) a.u. for
the Coulomb-Green function calculation.
 This comparison  clearly demonstrates
a significant improvement in accuracy,
 if the DKB basis is employed instead of the old one.
More extensive calculations employing the DKB basis will
be presented in forthcoming papers.

 This work was supported in
part by RFBR (Grant No. 01-02-17248), by the program 
"Russian Universities"  (Grant No. UR.01.01.072), and
by the Russian Ministry of Education (Grant No. E02-3.1-49).
The work of V.M.S. was supported by the Alexander von 
Humboldt Stiftung. V.A.Y. acknowledges the support of
the foundation "Dynasty" and the International Center
for Fundamental Physics. G.P. and G.S. acknowledge
financial support by the BMBF, DFG, and GSI.

%\end{document}

%\newpage
%\onecolumn

\begin{table}
\caption{The partial-wave contributions to the many-potential term
(in a.u.) for the $1s$ state at $Z=20$,
obtained by the basis set methods and
by the Coulomb-Green function (CGF) method.
The number of the basis functions: $n=60$. 
The shell model for the nuclear charge distribution has been used
with $R=3.478$ fm.}
%\vspace{0.5cm}
\begin{tabular} {l|l|l|l}
Term &Old basis
                 & DKB basis
                                     & CGF method \\
\hline 
% -1 &  0.838386 & 0.838398 & 0.838403 & 0.838427 & 0.838419 & 0.838416 &
$|\kappa|=1$ &0.848691 & 0.848750  & 0.848741
\\ 
$|\kappa|=2$ & 0.020618 &  0.020662 & 0.020653
\\
$|\kappa|=3$ &0.005302 & 0.005331 & 0.005326
\\
$|\kappa|=4$ & 0.002121 & 0.002139 & 0.002137
\\
$|\kappa|=5$ & 0.001050 & 0.001062 & 0.001062
\\
$|\kappa|=6$ & 0.000590 & 0.000597 &  0.000598
\\
%$|\kappa|=7$ & 0.000361 & 0.000365 & 0.000367
%\\
%$|\kappa|=8$ & 0.000235 &  0.000237 & 0.000240
%\\
%$|\kappa|=9$ & 0.000160 & 0.000161 & 0.000164
%\\
$|\kappa|\le$ 9 &  0.879127 & 0.879303 & 0.879288
\\
$|\kappa|\ge$ 10 & 0.000587 & 0.000585 & 0.000583
\\
Tot. many-pot. & 0.87971 & 0.87989 & 0.87987(1)
\\
\hline
\end{tabular}
\end{table}

\end{document}